\documentclass[twocolumn,amsmath,amssymb,prd]{revtex4}

\def\al{\alpha}
\def\be{\beta}
\def\ga{\gamma}
\def\de{\delta}
\def\ep{\epsilon}

\def\et{\eta}

\def\ka{\kappa}
\def\la{\lambda}

\def\rh{\rho}

\def\si{\sigma}

\def\ta{\tau}

\def\De{\Delta}

\def\Ph{\Phi}

\def\cl{{\cal L}}
\def\fr#1#2{{{#1} \over {#2}}}
\def\prt{\partial}
\def\pt#1{\phantom{#1}}
\def\half{{\textstyle{1\over 2}}}
\def\frac#1#2{{\textstyle{{#1}\over {#2}}}}

\def\lsim{\mathrel{\rlap{\lower4pt\hbox{\hskip1pt$\sim$}}
    \raise1pt\hbox{$<$}}}
\def\gsim{\mathrel{\rlap{\lower4pt\hbox{\hskip1pt$\sim$}}
    \raise1pt\hbox{$>$}}}

\def\curl#1{\vec\nabla\times\vec #1}
\def\div#1{\vec\nabla\cdot\vec #1}
\def\grad#1{\vec\nabla #1}
\def\etal {{\it et al.}}

\newcommand{\beq}{\begin{equation}}
\newcommand{\eeq}{\end{equation}}
\newcommand{\bea}{\begin{eqnarray}}
\newcommand{\eea}{\end{eqnarray}}
\newcommand{\rf}[1]{(\ref{#1})}

\def\kaf{k_{AF}}
\def\kf{k_{F}}
\def\kft#1#2#3{(k_{F})^{#1}_{\pt{#1}{#2}{#1}{#3}}} 
\def\kfi{(k_{F})_{\ka\la\mu\nu}}
\def\kt{\tilde k}
\def\ko{\tilde \ka_{o+}}
\def\ke{\tilde \ka_{e-}}
\def\ktr{\tilde \ka_{\rm tr}}

\begin{document}

\title{Lorentz-Violating Electrostatics and Magnetostatics}
\author{Quentin G.\ Bailey and V.\ Alan Kosteleck\'y}
\affiliation{Physics Department, Indiana University,
Bloomington, IN 47405, U.S.A.}
\date{IUHET 472, July 2004}

\begin{abstract}
The static limit of Lorentz-violating electrodynamics
in vacuum and in media is investigated.
Features of the general solutions include
the need for unconventional boundary conditions
and the mixing of electrostatic and magnetostatic effects.  
Explicit solutions are provided for some simple cases.
Electromagnetostatics experiments show promise 
for improving existing sensitivities 
to parity-odd coefficients for Lorentz violation
in the photon sector.
\end{abstract}

\maketitle


\section{Introduction}
\label{Introduction}

Since its inception,
relativity and its underlying Lorentz symmetry 
have been intimately linked to classical electrodynamics.
A century after Einstein,
high-sensitivity experiments based on electromagnetic phenomena
remain popular as tests of relativity.
At present,
many of these experiments are focused
on the ongoing search for minuscule violations of Lorentz invariance 
that might arise in the context of an underlying unified theory 
at the Planck scale
\cite{cpt01}.

Much of the work involving relativity tests with electrodynamics
has focused on the properties of electromagnetic waves,
either in the form of radiation or in resonant cavities.
Modern versions of 
the Michelson-Morley and Kennedy-Thorndike experiments
using resonant cavities
are among the best laboratory tests for relativity violations
\cite{cavexpt1,cavexpt2,cavexpt3},
while spectropolarimetric studies of cosmological birefringence
currently offer the most sensitive measures of Lorentz symmetry
in any system
\cite{photonexpt,kmphot}.
However,
the presence of Lorentz violation in nature would 
also affect other aspects of electrodynamics.
Our primary goal in this work is to initiate the
study of Lorentz-violating effects
in electrostatics and magnetostatics.
We find a variety of intriguing effects,
the more striking of which may make feasible 
novel experimental tests attaining exceptional sensitivities 
to certain types of relativity violations.

The analysis in this work is performed within the framework
of the Standard-Model Extension (SME)
\cite{ck,ak},
which enlarges general relativity and the Standard Model (SM) 
to include small arbitrary violations of Lorentz and CPT symmetry.
The full lagrangian of the SME can be viewed 
as an effective field theory for gravitational and SM fields
that incorporates all terms invariant 
under observer general coordinate and local Lorentz transformations.
Terms having coupling coefficients with Lorentz indices
control the Lorentz violation,
and they could emerge as low-energy remnants 
of the underlying physics at the Planck scale
\cite{kps}.
Experimental tests of the SME performed to date include
ones with
photons
\cite{cavexpt1,cavexpt2,cavexpt3,photonexpt,kmphot},
electrons
\cite{eexpt,eexpt2,eexpt3},
protons and neutrons
\cite{ccexpt,spaceexpt},
mesons
\cite{hadronexpt},
and muons
\cite{muexpt},
while interesting possibilities exists for 
neutrinos
\cite{nu}
and the Higgs
\cite{higgs}.

In the present work,
we limit attention to the sector of the minimal SME comprising
classical Lorentz-violating electrodynamics in Minkowski spacetime,
coupled to an arbitrary 4-current source.
There exists a substantial theoretical literature
discussing the electrodynamics limit of the SME 
\cite{photonth},
but the stationary limit remains unexplored to date. 
We begin by providing some general information about this theory,
including some aspects associated with macroscopic media.
We then adapt Green-function techniques 
to obtain the general solution for the 4-potential 
in the electrostatics and magnetostatics limit.
Among the associated unconventional effects
is a mixing of electric and magnetic phenomena
that is characteristic of Lorentz violation. 
For field configurations in Lorentz-violating electromagnetostatics,
we show that the usual Dirichlet or Neumann boundary conditions
are replaced by four natural classes of boundary conditions,
a result also reflecting this mixing.

As one application,
we obtain the Lorentz-violating 4-potential
due to a stationary point charge.
The usual radial electrostatic field is corrected by
small Lorentz-violating terms,
and a small nonzero magnetostatic field also emerges.
Another solution presented here describes 
a nonzero scalar potential arising inside a conducting shell 
due to a purely magnetostatic source placed within the shell.
This configuration appears well suited as the basis 
for a high-sensitivity experiment
that would seek certain nonzero parity-breaking effects 
in Lorentz-violating electrodynamics.
We discuss some aspects of an idealized experiment of this type,
including the use of rotations and boosts to extract the signal.
Finally, 
the appendix resolves some basic issues associated 
with coordinate redefinitions in the illustrative case 
of a classical charged point particle. 
Throughout this work,
we adopt the conventions 
of Refs.\ \cite{kmphot,ck}.

\section{Framework}
\label{Framework}

\subsection{Vacuum electrodynamics}
\label{Vacuum electrodynamics}

The lagrangian density for the photon sector of the minimal SME 
can be written as
\bea 
\cl &=& -\frac 1 4 F_{\mu\nu}F^{\mu\nu}
- \frac 1 4 \kfi F^{\ka\la}F^{\mu\nu}
\nonumber\\ & & 
+\frac 1 2 (\kaf)^\ka\ep_{\ka\la\mu\nu}A^\la F^{\mu\nu}
- j^\mu A_\mu. 
\label{L_em1}
\eea
In this equation,
$j^\mu = (\rh, \vec J)$ is the 4-vector current source
that couples to the electromagnetic 4-potential $A_\mu$,
and $F_{\mu\nu} \equiv \prt_\mu A_\nu - \prt_\nu A_\mu$
is the electromagnetic field strength,
which satisfies the homogeneous equations 
\beq
\ep^{\mu\nu\ka\la}\prt_\mu F_{\ka\la}=0 
\label{homeqns}
\eeq
ensuring the U(1) gauge invariance of the action.
The coefficients $\kfi$ and $(\kaf)^\ka$
control the Lorentz violation and are expected to be small.

To focus the analysis,
some simplifying assumptions are adopted in what follows.
Since our primary interest here is electromagnetostatics,
we take the current $j_\mu$ to be conventional.
Lorentz violation is then present only in the photon sector,
and the Lorentz force is conventional.
This limit is less restrictive than might first appear,
since suitable coordinate redefinitions can
move some of the Lorentz-violating effects into the matter sector
without changing the physics. 
An explicit discussion of this issue for the case of a point charge
is given in the appendix.

In the simplest scenarios for Lorentz violation
the coefficients $\kfi$ and $(\kaf)^\ka$ are constant,
so that energy and momentum are conserved.
We adopt this assumption here.
Variation of the lagrangian \rf{L_em1}
then yields the inhomogeneous equations of motion
\beq
\prt_\al{F_\mu}^\al
+(\kf)_{\mu\al\be\ga}\prt^\al F^{\be\ga}
+ (k_{AF})^\al \ep_{\mu\al\be\ga} F^{\be\ga}
+j_{\mu}=0, 
\label{eqmotphot1}
\eeq
which extend the usual covariant Maxwell equations 
to incorporate Lorentz violation.
Although outside our present scope,
a treatment allowing nonconservation of energy-momentum 
would be of interest. 

Some of the analysis of this theory is simplified 
by introducing certain convenient linear combinations of 
the coefficients $\kfi$ for Lorentz violation.
One useful set is given by
\cite{kmphot} 
\bea
(\ka_{DE})^{jk} &=& -2 (\kf)^{0j0k}, 
\nonumber \\
(\ka_{HB})^{jk} &=& \half \ep^{jpq} \ep^{krs} (\kf)^{pqrs}, 
\nonumber \\
(\ka_{DB})^{jk} &=& -(\ka_{HE})^{kj} = (\kf)^{0jpq}\ep^{kpq}.
\label{kappas}
\eea
As an immediate application of these definitions,
the microscopic equations \rf{eqmotphot1}
for Lorentz-violating electrodynamics \it in vacuo \rm
with $(k_{AF})^\ka = 0$
can be cast in the form of the Maxwell equations
for macroscopic media,
\bea
\div D = \rh ,
& &
\quad 
\curl H - \prt_0 {\vec D} = \vec J , 
\nonumber \\
\div B = 0 ,
& &
\quad 
\curl E + \prt_0 {\vec B} = 0 ,
\label{eqmotphot2}
\eea
by adopting the definitions 
\bea
\vec D &\equiv& (1 + \ka_{DE})\cdot \vec E  + \ka_{DB} \cdot \vec B,
\nonumber\\
\vec H &\equiv& (1 + \ka_{HB})\cdot \vec B  + \ka_{HE} \cdot \vec E, 
\label{vacuum}
\eea
which hold in vacuum.

Another useful set of combinations is 
\cite{kmphot}
\bea
(\tilde\ka_{e+})^{jk}&=&\half(\ka_{DE}+\ka_{HB})^{jk},
\nonumber\\
(\tilde\ka_{e-})^{jk}&=&\half(\ka_{DE}-\ka_{HB})^{jk}
                 -\frac13\de^{jk}(\ka_{DE})^{ll},
\nonumber\\
(\tilde\ka_{o+})^{jk}&=&\half(\ka_{DB}+\ka_{HE})^{jk} ,
\nonumber\\
(\tilde\ka_{o-})^{jk}&=&\half(\ka_{DB}-\ka_{HE})^{jk},
\nonumber\\
\tilde\ka_{\rm tr}&=&\frac13(\ka_{DE})^{ll}.
\label{kappas2}
\eea
This set is of particular relevance for 
certain experimental considerations.
For example,
if $(k_{AF})^\ka = 0$
then birefringence induced by Lorentz violation
is controlled entirely by the 10 coefficients  
$(\tilde\ka_{e+})^{jk}$ and $(\tilde\ka_{o-})^{jk}$.

Experiments constrain part of the space 
of coefficients for Lorentz violation
\cite{kmphot,photonexpt}.
The CPT-odd coefficients $(\kaf)^{\ka}$ are stringently bounded 
by cosmological observations
and are set to zero throughout this work.  
Spectropolarimetry of cosmologically distant sources 
limits the 10 combinations
$(\tilde\ka_{e+})^{jk}$ and $(\tilde\ka_{o-})^{jk}$
to values below parts in $10^{32}$.
The remaining 9 linearly independent components of $\kfi$
are accessible in laboratory experiments
\cite{cavexpt1,cavexpt2,cavexpt3},
with the best sensitivity achieved to date
on some components of 
$(\tilde\ka_{e-})^{jk}$ 
at the level of about $10^{-15}$.

For some of what follows,
it is useful to consider the limit of the theory \rf{L_em1}
in which the 10 coefficients  
$(\tilde\ka_{e+})^{jk}$ and $(\tilde\ka_{o-})^{jk}$
are set identically to zero.
The pure-photon part of the lagrangian \rf{L_em1} then 
can be written as
\beq 
\cl = -\frac 1 4 F^{\ka\la}F^{\mu\nu}
[ \et_{\ka\mu} \et_{\la\nu} 
+ \et_{\ka\mu} \kft\al\la\nu 
+ \et_{\la\nu} \kft\al\ka\mu ].
\\
\label{L_em2}
\eeq
In this limit,
corresponding simplifications occur
in the combinations \rf{kappas}.
For example,
$(\ka_{DB})^{jk}$
becomes an antisymmetric matrix.

\subsection{Electrodynamics in media}
\label {Electrodynamics in media}

The analysis of electrodynamics in ponderable media
could in principle proceed
by supplying a four-current density that describes 
the microscopic charge and current distributions in detail.
However,
as usual,
it is more practical to adopt an averaging process.  

Following standard techniques 
\cite{jdj},
we average the exact Maxwell equations \rf{eqmotphot2} 
with the microscopic charge and current densities
over elemental spatial regions.
We introduce the usual notions of
polarization $\vec P$ and magnetization $\vec M$ 
in terms of averaged molecular electric and magnetic dipole moments,
\bea
\vec P &=& \left< \sum_{n} \vec p_n \, 
\de^3 (\vec x - \vec x_n) \right>, 
\label{Pmoment} \nonumber\\
\vec M &=& \left< \sum_{n} \vec m_n \, 
\de^3 (\vec x - \vec x_n) \right>.
\label{Mmoment}
\eea
The braces here denote a smooth spatial average over many molecules,
and the sums are over each type $n$ of molecule 
in the given material.   
For present purposes these multipoles suffice,
and the higher-order microscopic moments 
arising from the averaging process can be ignored.

With these definitions,
the analogue displacement field $\vec D$ 
and analogue magnetic field $\vec H $ 
appearing in the vacuum equations \rf{eqmotphot2}
are replaced with the macroscopic displacement field
$\vec D_{\rm matter}$
and macroscopic magnetic field 
$\vec H_{\rm matter}$,
defined by 
\bea
\vec D_{\rm matter}
&=& (1 + \ka_{DE})\cdot \vec E  + \ka_{DB} \cdot \vec B 
+ \vec P, 
\nonumber\\
\vec H_{\rm matter}
&=& (1 + \ka_{HB})\cdot \vec B  + \ka_{HE} \cdot \vec E 
- \vec M. 
\label{dfield} 
\eea
The inhomogeneous Maxwell equations in macroscopic media
in the presence of Lorentz violation
then still take the form \rf{eqmotphot2}.   

The above formulation shows that,
unlike the Lorentz-symmetric case,
a scalar dielectric permittivity 
and a scalar magnetic permeability alone
may be insufficient to describe the linear
response of a simple material 
to applied electric and magnetic fields.
Instead,
we must allow for the existence of components of the 
induced moments that are orthogonal to the applied field,  
induced by atomic-structure modifications from the Lorentz violation.
The nature of the response may vary for different materials.
For homogeneous materials,
the unconventional response 
can be described by the constituency relations
\bea
\vec D_{\rm matter} &=& (\ep + \ka_{DE}^{\rm vacuum} 
+ \ka_{DE}^{\rm matter} )\cdot \vec E  
\nonumber\\
& & + (\ka_{DB}^{\rm vacuum} 
+ \ka^{\rm matter}_{DB} ) \cdot \vec B, 
\nonumber\\
\vec H_{\rm matter} &=& ( \frac 1 {\mu}  
+ \ka_{HB}^{\rm vacuum} + \ka^{\rm matter}_{HB} )\cdot \vec B 
\nonumber\\ 
& & + ( \ka_{HE}^{\rm vacuum} 
+ \ka^{\rm matter}_{HE} ) \cdot \vec E.  
\label{hfield2} 
\eea
Here,
the permittivity $\ep$ and the permeability $\mu$ 
are understood to be those in the absence of Lorentz violation,
the coefficients $\ka^{\rm vacuum}$ are those 
discussed in the previous subsection,
and the coefficients $\ka^{\rm matter}$ 
contain the pieces of the induced moments $\vec P$ and $\vec M$ 
that are leading-order in $\kfi$
and that may be partially or wholly orthogonal 
to the applied fields.   

The explicit form of the coefficients $\ka^{\rm matter}$ 
depends on the macroscopic medium.
Indeed, 
unless the matter is isotropic,
their values can depend on the orientation of the material.  
Applying the averaging process
to an appropriate atomic or molecular model based on the SME
could establish their form
and represents an interesting open problem. 
Note that, 
in analyzing an experiment,
it may be insufficient merely to replace
expressions involving vacuum coefficients
with ones involving the sum of vacuum and matter coefficients
because the boundary conditions in the presence of matter
may induce further modifications. 
An explicit example of this is given in subsection
\ref{Magnet in conducting shell}.

In the remainder of the paper, 
coefficients without labels
are understood to be those in the vacuum.   
The relevant matter coefficients are explicitly labeled
as $\ka^{\rm matter}$.

\section{Electromagnetostatics}
\label{Electromagnetostatics}

In this section,
we consider stationary solutions of the theory \rf{eqmotphot1}.
The stationary fields \it in vacuo \rm  
satisfy the time-independent equation of motion 
\beq
\kt^{j\mu k\nu} \prt_j \prt_k
A_{\nu}(\vec x) = j^{\mu}(\vec x).
\label{eqmotphot3} 
\eeq
In this equation,
the coefficients $\kt^{j\mu k\nu}$ are defined by
\beq 
\kt^{j\mu k\nu} = 
\et^{jk}\et^{\mu\nu}-\et^{\mu k} \et^{\nu j}+2(\kf)^{j\mu k\nu}. 
\label{diffop}
\eeq
From Eq.\ \rf{homeqns},
the electrostatic and magnetostatic fields 
can be written in terms of 
a 4-potential $A^\la = (\Ph, A^j)$ according to
$\vec E = - \grad \Ph$ and $\vec B = \curl A$,
as usual.

The appearance of both $\vec E$ and $\vec B$
in $\vec D$ and in $\vec H$,
which occurs even in the vacuum (cf.\ Eq.\ \rf{vacuum}),
means that in the presence of Lorentz violation
a static charge density generates both an electrostatic 
and a (suppressed) magnetostatic field,
and similarly
a steady-state current density generates both a magnetostatic 
and a (suppressed) electrostatic field.
We show below that the subjects of electrostatics and magnetostatics,
which are distinct in the usual case,
become convoluted in the presence of Lorentz violation.
A satisfactory discussion therefore requires 
the simultaneous treatment of both electric and magnetic phenomena,
even in the static limit
\cite{fn1}.

\subsection{Green functions and boundary conditions}
\label{Green functions and boundary conditions}

To obtain a general solution for the potentials $\Ph$ and $\vec A$,
we introduce indexed Green functions
$G_{\mu\al}(\vec x,\vec x')$ 
solving Eq.\ \rf{eqmotphot3} for a point source,  
\beq
\kt^{j\mu k\nu} \prt_j \prt_k G_{\mu\al}(\vec x,\vec x')  
= \de^{\nu}_{\pt{\nu}\al}\de^3(\vec x -\vec x').
\label{eqgreen}
\eeq
Once a suitable Green theorem 
incorporating the differential operator in Eq.\ \rf{eqmotphot3} 
is found,  
the formal solution of Eq.\ \rf{eqmotphot3}
can be constructed using standard methods
\cite{jdj,mf}.

The relevant Green theorem can be given 
in terms of arbitrary functions 
$X_{\mu}(\vec x)$ and $Y_{\mu}(\vec x)$.
We obtain
\bea
&
\hskip-20pt
\int_V d^3x ( 
X_\mu \kt^{j\mu k\nu} \prt_j \prt_k Y_\nu
-Y_\nu \kt^{j\mu k\nu} \prt_j \prt_k X_\mu
)
\nonumber\\
& =
-\int_S d^2S ~ \hat n^j (
Y_\mu \kt^{j\mu k\nu} \prt_k X_\nu
-X_\mu \kt^{j\mu k\nu} \prt_k Y_\nu
), 
\label{grthm2}
\eea
where $\hat n^j$ is the outward normal of the surface $S$ 
bounding the region of interest.  

Using Eqs.\ \rf{eqmotphot3} and \rf{eqgreen} 
in Eq.\ \rf{grthm2},
we find that the general solution 
for the 4-potential $A^\la$ is
\bea
A_{\la}(\vec x) &=& 
\int_V d^3x' G_{\mu\la}(\vec x',\vec x) j^{\mu}(\vec x')  
\nonumber\\ 
& - & 
\int_S d^2S' \hat n'^j [ 
G_{\mu\la}(\vec x',\vec x) \kt^{j\mu k\nu}\prt'_k A_{\nu}(\vec x') 
\nonumber\\ & & 
\qquad \qquad
- A_{\mu}(\vec x') \kt^{j\mu k\nu}\prt'_k G_{\nu\la} (\vec x',\vec x)
]
\label{potsol}
\eea
up to the gradient of an arbitrary scalar.
The first term contains the contribution from the sources 
within the volume $V$,
while the remainder represents 
the effects of the bounding surface $S$.

Consider next the issue of appropriate boundary conditions.
Inspection of Eq.\ \rf{potsol} reveals that  
there are four natural classes of boundary condition
that specify a solution.
We summarize these in Table 1.

\begin{center}
\begin{tabular}{|c|c|c|}
\hline
\multicolumn{1}{|c|}{ } & 
\multicolumn{1}{|c|}{ Fields on $S$ } & 
\multicolumn{1}{c|}{ Green function on $S$ }
\\
\hline
I &
$\Ph$, $\hat n \times \vec A$ 
& $ \ep^{jkl} \hat n^j G^l_{\pt{l}\la}=0$,
$G_{0\la} = 0$ 
\\ 
II &
$\Ph$, 
$\hat n \times \vec H $ 
& $ \hat n^j \kt^{j l k\nu}\prt_k G_{\nu\la}
=\frac 1 S\de^l_{\pt{l}\la}$,
$G_{0\la} = 0$  
\\ 
III &
$\hat n \cdot \vec D $, $\hat n \times \vec A$
& $ \hat n^j \kt^{j 0 k\nu}\prt_k G_{\nu\la}
=\frac 1 S\de^0_{\pt{0}\la}$, 
$\ep^{jkl} \hat n^j G^l_{\pt{l}\la}=0$ 
\\
IV &
$\hat n \cdot \vec D$, 
$\hat n \times \vec H $
& $\hat n^j \kt^{j\mu k\nu}\prt_k G_{\nu\la}
=\frac 1 S\de^\mu_{\pt{\mu}\la}$
\\
\hline
\end{tabular}
\end{center}
\begin{center}
Table 1. Natural classes of boundary conditions.
\label{const}
\end{center}
  
Class I boundary conditions are expressed entirely 
in terms of the potentials,
being specified by $\Ph\equiv A^0$
and the tangential component of $\vec A$.
This class is the only one for which the explicit
Green function is independent of the area of the surface.
It is most closely analogous to Dirichlet boundary conditions
in conventional electrostatics.
Class II boundary conditions involve $\Ph$
and the tangential component of $\vec H$.
The explicit boundary condition on the corresponding Green function 
incorporates a factor of the inverse surface area 
on the right-hand side,
generating a term in the solution involving 
the average contribution of the potential
over the bounding surface.
A similar feature occurs in conventional electrostatics
for Neumann boundary conditions.
Class III boundary conditions involve 
the normal component of $\vec D$ 
and the tangential component of $\vec A$,
while class IV boundary conditions involve 
the normal component of $\vec D$ 
and the tangential component of $\vec H$.
The gauge freedom in specifying $\vec A$
in class I and III boundary conditions
has no effect on the solutions for $\vec E$ and $\vec B$.

The uniqueness of the solutions for the fields
$\vec E$ and $\vec B$ associated to each of the four classes above
can be shown by using the general solution \rf{potsol}  
to examine the difference 
$\De A_\mu = A_\mu^1 - A_\mu^2$ 
of two solutions obtained 
for a specified choice of boundary conditions.
Direct calculation verifies that $\De A_\mu$ 
is either a constant or zero.  
It follows that the electric and magnetic fields 
from both solutions are identical.
This result can also be verified without 
the use of Green functions 
by considering the first Green identity 
and the field boundary conditions.

We remark in passing that reciprocity relations 
for the Green functions can be obtained 
by combining the Green theorem \rf{grthm2} with the above results.
For example,
provided the bounding surface is at infinity,
the Green functions satisfy 
$G_{\mu \nu} (\vec x, \vec x') = G_{\nu \mu} (\vec x', \vec x)$.

Note also that a given problem 
in Lorentz-violating electromagnetostatics
effectively requires the simultaneous solution 
of all four potentials $A_{\rh}$
from the corresponding boundary conditions.
This differs from the usual case,
in which electrostatics and magnetostatics 
can be regarded as distinct subjects,
even though Eq.\ \rf{potsol} reduces to standard results 
in the limit of zero $\kf$.

The solution \rf{potsol} can be generalized 
to include regions of matter.  
In such cases the modified equations \rf{hfield2} apply.  
For simplicity,
we limit attention to cases where the material is isotropic
and the matter coefficients are constant everywhere inside.
The equations of motion for the electromagnetic field are then
\bea 
\kt^{j\mu k\nu}_{\rm matter} \prt_j \prt_k
A_{\nu}(\vec x) = j^{\mu}(\vec x),
\label{eqmotphotm}
\eea
where the coefficients $\kt^{j\mu k\nu}_{\rm matter}$ are given by
\bea
\kt^{j\mu k\nu}_{\rm matter} &=& 
\ep \et^{jk} \et^{\mu 0} \et^{\nu 0}
- \frac 1 \mu \et^{jk} \et^{\mu l} \et^{\nu l} 
- \frac 1 \mu \et^{j \nu} \et^{k \mu} 
\nonumber\\  && 
+ 2 (\kf)^{j\mu k\nu}_{\rm vacuum}
+ 2(\kf)^{j\mu k\nu}_{\rm matter}. 
\label{diffop2}
\eea
In this equation,
the coefficients $(\kf)_{\rm matter}$ 
are related to the quantities $\ka^{\rm matter}$
in Eq.\ \rf{hfield2}
by definitions of the form \rf{kappas}.
Assuming that $(\kf)_{\rm matter}$ 
has the symmetries of $(\kf)_{\rm vacuum}$ 
and that the volume $V$ is filled with matter to the surface $S$,
the solution \rf{potsol} and the above formalism 
can be applied directly by replacing
$\kt^{j\mu k\nu}_{\rm vacuum} \to \kt^{j\mu k\nu}_{\rm matter}$.

\subsection{Conductors}
\label{Conductors}

The presence of conducting surfaces 
influences the determination of appropriate boundary conditions.
In conventional electrostatics, 
the potential $\Ph$ is constant 
on a conductor in equilibrium.
However,
it unclear \it a priori \rm whether this result
holds in the presence of Lorentz violation,
when the potential $\Ph \equiv \Ph_\rh + \Ph_J$
becomes the sum of a part $\Ph_\rh$ arising from the charge density
and a (suppressed) part $\Ph_J$ arising from the current density.

To investigate this issue,
consider one or more conductors positioned in a region 
that may also contain static charges and steady-state currents.
Assuming the region contains matter
with the general constituency relations \rf{hfield2},
the electromagnetic energy density $u$ can be written 
\bea
u &=& 
\half [ 
\vec E \cdot ( \ep + \ka_{DE}^{\rm vacuum} 
+ \ka_{DE}^{\rm matter} )\cdot \vec E 
\nonumber\\
&& 
\quad
+ \vec B \cdot (\frac 1 \mu + \ka_{HB}^{\rm vacuum} 
+ \ka_{HB}^{\rm matter}) \cdot \vec B
],
\label{energy1}
\eea
where for simplicity $\ka^{\rm matter}$ and $\ka^{\rm vacuum}$
are taken to have the same symmetries.
The total electromagnetic energy $U$ of the configuration
is the integral of Eq.\ \rf{energy1} over all space, 
including the volume occupied by the conductors.  
If the fields fall off sufficiently rapidly 
at the boundary of the region, 
the total energy can alternatively be expressed 
as an integral over all space involving
the potentials $\Ph$ and $\vec A$, 
the charge density $\rh$,
and the current density $\vec J$.

In equilibrium,
the free charge on the conductor is arranged 
to minimize $U$.
Consider a variation $\de\rh$ 
of the charge distribution on the conductors 
away from the equilibrium configuration.
With some manipulation of the above expressions
and suitable use of the modified Maxwell equations,
the corresponding change $\De U$ in the electromagnetic energy 
can be written as   
\bea
\De U &\equiv& U(\rh +\de \rh) - U(\rh)
\nonumber\\
&=&
\int_V d^3x [ 
\Ph_{\rh}  \de \rh 
\nonumber\\
&& 
\quad
+ \half \de \vec E \cdot (\ep  
+ \ka_{DE}^{\rm vacuum} 
+ \ka_{DE}^{\rm matter}) \cdot \de \vec E 
\nonumber\\
&& \quad
+\half \de \vec B \cdot (\frac 1 \mu
+ \ka_{HB}^{\rm vacuum} 
+ \ka_{HB}^{\rm matter}) \cdot \de \vec B 
].  
\label{envar2}
\eea
The first term represents a first-order variation,
so the energy is extremized when it vanishes.
This condition is satisfied for constant $\Ph_{\rh}$ 
in the conductors because the variation $\de\rh$ integrates to zero.   
The remaining terms represent the second-order variation, 
and the energy is minimized when they are positive.
This condition is also satisfied
because $\ka_{DE}$ and $\ka_{HB}$ are small.
The potential $\Ph_\rh$ from the charge density 
is thus expected to be constant in a conductor in equilibrium.
This result generalizes the Thomson theorem 
of conventional electrostatics, 
and it establishes a partial condition on $\Ph$.

The energy-based variational approach of Eq.\ \rf{envar2} 
gives no information about the portion $\Ph_J$ of the potential 
due to the current density.  
This is because 
the contribution of $\Ph_J$ to $\Ph$ in the expression for the energy 
cancels against the portion $\vec A_\rh$ of $\vec A$
arising from the charge density.
Instead,
a condition on $\Ph_J$
can be obtained by considering the rate of work done 
by the fields in a fixed volume of conductor.

Since the Lorentz force is conventional,
the power $P$ in the volume $V$ of conductor is given by
\beq
P = -\int_V d^3x ~ \vec J \cdot \vec E. 
\label{power}
\eeq
Using $\vec E = -\grad \Ph$, the steady-state assumption
$\div J = 0$, and the vanishing of the normal component
of the current at the surface of the conductor 
reveals that $P$ vanishes.
Substituting the steady-state Maxwell equation
$\curl H_{\rm matter} = \vec J$ in Eq.\ \rf{power}
and using $\curl E = 0$ then gives the condition 
\beq
\int_S d^2x ~ \vec E \cdot \hat n \times \vec H_{\rm matter} = 0
\label{surfaceflow}
\eeq
on the surface of the conductor.
This is the statement that there is no net outward flow 
of field momentum:
the integral of the normal component 
of the generalized Poynting vector
$\vec S =  \vec E \times \vec H_{\rm matter}$ 
vanishes over the surface.
The point of interest is that Eq.\ \rf{surfaceflow} 
is satisfied for vanishing tangential electric field 
at the surface,
which in turn implies that $\Ph_J$ is constant.
The condition \rf{surfaceflow} is therefore consistent 
with requiring that the total potential $\Ph$ 
is constant on the surface of a conductor.  

Note that Eq.\ \rf{surfaceflow} could in principle 
also be satisfied for more general field configurations. 
Suppose that for linear conductors
we introduce a generalized Ohm law of the form
\beq
\vec J = (\si + \tilde \si_E) \cdot \vec E 
+ \tilde \si_B \cdot \vec B,
\label{ohmslaw}
\eeq
where $\tilde \si_E$, $\tilde \si_B$
involve coefficients for Lorentz violation.
It can be shown using the above assumptions 
that only the final term in this expression
could produce leading-order deviations from constant $\Ph$ 
in equilibrium.
A term of this type might conceivably be generated 
through Lorentz violation under suitable circumstances.
Although outside our present scope,
it would be of interest to investigate this issue
within specific models of conductors.

As an aside,
we remark that the above results can be used to show that 
certain precision Cavendish-type experiments 
searching for a nonzero photon mass $\mu$ 
are insensitive to Lorentz violation.
For example,
the value of $\mu$ in the Proca electrostatics equation
\cite{photmassrev}
$(\vec \nabla^2 - \mu^2)\Ph = 0$ 
was bounded by Williams \etal\ \cite{photmass}.
This experiment basically sought a nonzero potential difference 
inside a metal shell held at fixed potential.   
However,
at leading order,
the corresponding equation with a nonzero $\kf$ coefficient 
for Lorentz violation is the Laplace equation.
Fixing the potential across the shell 
thus also fixes it inside the shell,
and so the experiment is insensitive to $\kf$.

\section{Applications}
\label{Applications}

\subsection{Point charge}   
\label{Point charge}   

As an application of the general solution 
using the Green function \rf{potsol},
consider the fields for the special case 
of boundary conditions at infinity.
The potentials can then be obtained from 
\bea
A_{\la}(\vec x) & = & \int d^3x ~ 
G_{\mu\la}(\vec x, \vec x') j^{\mu} (\vec x').
\label{a}
\eea
Imposing the Coulomb gauge,
this integral can be solved at leading order in $\kfi$
for an arbitrary source $j^{\mu}$,
using the symmetric Green function
\bea
G_{\mu\nu} (\vec x - \vec x') & = & 
{ \et_{\mu\nu} + (\kf)_{\mu j \nu j} \over 4 \pi |\vec x - \vec x'| }
\nonumber\\
&&
- { (\kf)_{\mu j \nu k} (\vec x -\vec x')^j (\vec x -\vec x')^k
\over 4 \pi |\vec x - \vec x'|^3 }.
\label{b}
\eea
This Green function can be extracted 
from the differential equations \rf{eqmotphot3} and \rf{eqgreen} 
by Fourier decomposition in momentum space.

As an example,
consider a classical point charge
as the source.
The action for this case is discussed in the appendix,
along with some of the subtleties associated with 
the freedom to redefine the choice of coordinates.
In the rest frame of the charge,
taken to be located at the origin,
the source 4-current is
$j^\mu (\vec x) = \de_0^{\pt{0}\mu} q \de^{(3)}(\vec x)$.
Substituting this source and the Green function \rf{b}
into the solution \rf{a} 
and performing the integral,
we find the potentials at leading order in Lorentz violation
are given by 
\bea
\Ph (\vec x) &=&  \fr q {4\pi|\vec x|}
\Big( 
1 - (\kf)^{0j0k} \hat x^j \hat x^k
\Big) ,
\nonumber\\
A^j(\vec x) &=& \fr q {4\pi |\vec x|} 
\Big( 
(\kf)^{0kjk} - (\kf)^{jk0l} \hat x^k \hat x^l 
\Big).
\label{c}
\eea
We have defined the charge $q$ so that the first term
in $\Ph$ has the usual normalization. 
The solution for $\Ph$ agrees with that 
previously obtained in Ref.\ \cite{kmphot}.
As discussed above,
the appearance of a nonzero vector potential
from a point charge at rest is to be expected 
and can be traced to the mixing 
of electrostatics and magnetostatics
in the presence of Lorentz violation.

The electromagnetostatic fields due to the point charge at rest
can be derived directly from the results \rf{c}.
We find
\bea 
E^j(\vec x) & = & 
\fr {q} {4\pi |\vec x|^2}
\Big(
\hat x^j
+ {2(\kf)^{0j0k} \hat x^k}
\nonumber\\
&& 
\qquad\qquad
- 3(\kf)^{0k0l} \hat x^j \hat x^k \hat x^l
\Big), 
\nonumber\\
B^j(\vec x) & = & 
\fr {q } {4\pi |\vec x|^2}\ep^{jkl} 
\Big((\kf)^{0mkm} \hat x^l
\nonumber\\
&& 
\qquad\qquad
+[ (\kf)^{0kml} + (\kf)^{0mkl} ]\hat x^m
\nonumber\\
&& 
\qquad\qquad
+ 3 (\kf)^{0mnk} \hat x^m \hat x^n \hat x^l
\Big).
\label{d}
\eea
These fields display an inverse-square behavior
modulated by anisotropic Lorentz-violating parts.

\subsection{Magnet inside conducting shell}
\label{Magnet in conducting shell}

We consider next a more involved example,
consisting of a localized magnetic source 
surrounded by a grounded conducting shell. 
This situation is designed to exploit the 
mixing between electrostatic and magnetostatic effects 
in a manner that has direct application 
to laboratory searches for Lorentz violation,
as is discussed in the next subsection.

For definiteness,
the magnetic source is taken to be
a sphere of radius $a$ and uniform magnetization $\vec M$.
At zeroth order in Lorentz violation,
the associated magnetic field is uniform inside the sphere 
and is dipolar outside, 
with dipole moment $\vec m =  4\pi a^3\vec M/3$.   
The grounded conductor 
is taken to be a concentric spherical shell of radius $R$. 
We seek the solution for the scalar potential $\Ph$
in the region $a<r<R$,
where $r$ is the radial coordinate from the center of the sphere.
We first solve the problem treating the source
as an idealized bound current density, 
and then present the modifications
induced by the magnet permittivity and
matter coefficients for Lorentz violation. 

The idealized solution can be found using Eq.\ \rf{potsol} 
with class I boundary conditions,
\bea
\Ph(\vec x) &=& \int_V d^3x'
G_{j0}(\vec x',\vec x) j^{j}(\vec x')  
\nonumber\\ & &
+ \int_S d^2S' ~ \hat n'^j 
A_{\mu}(\vec x') \kt^{j\mu k\nu}\prt'_k G_{\nu 0} (\vec x',\vec x) .
\label{msol1}
\eea
The charge density $\rh$ vanishes by assumption,
so the source consists of the bound current density 
$\vec J = \curl M$
due to the magnetization of the sphere.
At leading order in the coefficients for Lorentz violation, 
Eq.\ \rf{msol1} can be manipulated into the form
\bea
\Ph^{(1)} (\vec x)= 
-\int_V d^3x'
G^{(0)}(\vec x,\vec x') 
\grad' \cdot \ka_{DB} \cdot \vec B^{(0)}(\vec x') .
\label{msol2}
\eea
The labels $(0)$ and $(1)$ indicate zeroth- and first-order 
contributions in the coefficients for Lorentz violation.  

The structure of Eq.\ \rf{msol2} implies that 
the potential can be viewed as arising 
from an effective charge density obtained from the
derivatives of the conventional magnetic field.
The proposed application of this problem lies in the laboratory,
so in evaluating Eq.\ \rf{msol2} it is an excellent approximation
to take $\ka_{DB} = \tilde\ka_{o+}$ as antisymmetric,
following the discussion in section \ref{Vacuum electrodynamics}.
For the zeroth-order Green function
$G^{(0)}(\vec x,\vec x')$,
we can take the conventional Dirichlet Green function 
for a spherical grounded shell of radius $R$.
Performing the integral,
we find that the Lorentz-violating potential $\Ph$ is 
\bea
\Ph (\vec x) = {\hat r \cdot \ko 
\cdot \vec m \over 4\pi } 
\left( {1 \over r^2} 
- {r \over R^3} \right). 
\label{finsoln}
\eea
This solution is valid for $a<r<R$.  

The electrostatic and magnetostatic fields 
can be obtained by direct calculation.
The electrostatic field in the region $a<r<R$ is given by
\bea
E^j (\vec x) & = & \fr{(\ko)^{jk} m^k}{4\pi}
\left( {1 \over R^3} - { 1 \over r^3} \right)
+ \fr {3(\ko)^{kl} \hat r^j \hat r^k m^l} {4\pi r^3}.
\nonumber \\
\label{e}
\eea
The magnetostatic field is given by
\bea
B^j (\vec x) & = & 
\fr {3\hat r^j (\hat r \cdot \vec m) - m^j}{4\pi r^3}
\label{f}
\eea
at zeroth order.

The solution \rf{finsoln} becomes modified 
in the more realistic scenario
with the magnet consisting of matter obeying the
constituency relations \rf{hfield2}.
To obtain the modified result
for the case of an isotropic material
with constant matter coefficients,
note that the leading-order potential $\Ph^{(1)}$
satisfies the Laplace equation everywhere except at $r=a$ 
and that the normal component of the electric field satisfies 
the boundary condition $\De (\ep E_n^{(1)}) = \si^{\rm eff}$,
where the effective surface charge $\si^{\rm eff}$ 
is determined by the discontinuity 
of the magnetic field at $r=a$.  
The problem is then formally identical 
to a conventional electrostatics problem,
and standard techniques 
\cite{jdj}
apply.
We find that the solution \rf{finsoln} 
is adjusted by the replacements
\bea
\vec m &\to& \fr{3\vec m}{[2+\ep + (1-\ep)a^3/R^3]},
\nonumber\\
\ko &\to& \ko^{\rm vacuum} + \frac 2 3 \ko^{\rm matter},
\eea
where the dielectric constant $\ep$ is taken to be a constant scalar.

\subsection{Experiment}
\label{Experiment}

Among the combinations of coefficients listed in Eq.\ \rf{kappas2},
experiments to date are least sensitive to $\ko$ and $\ktr$.
In the case of $\ko$,
this reduced sensitivity 
can be attributed to the parity-odd nature of the
corresponding Lorentz-violating effects,
while high sensitivity to $\ktr$ is difficult to attain  
because it is a scalar.
The configuration discussed in the previous subsection
is constructed to be directly sensitive to parity-odd effects,
as is reflected in the dominance of the combination $\ko$ 
in the solution \rf{finsoln}.

In this section,
we consider an idealized experiment that
could attain high sensitivity 
to the three independent components of $\ko$ 
and indirectly also to $\ktr$.
The idea is to measure the potential \rf{finsoln}
inside the spherical cavity.
For a conservative estimate of the sensitivity that
might in principle be attainable in the ideal case,
suppose for simplicity the spherical source 
is a hard ferromagnet 
with strength $10^{-1}$ T near its surface,
and suppose the potential is measured 
with a voltmeter of nV sensitivity.
Then,
a null measurement in principle could achieve a bound 
of $\ko \lsim 10^{-15}$,
which would represent an improvement of 
four orders of magnitude over best existing sensitivities
\cite{cavexpt2,cavexpt3}.
Using SQUID-based devices,
this might in principle be improved 
by another four orders of magnitude,
suggesting Planck-scale sensitivity to 
this type of Lorentz violation is attainable 
in the laboratory.

The basic setup for the experiment would be to  
insert one or more voltage probes, 
referenced to each other or to ground,
into the inner region $a<r<R$.  
The solution \rf{finsoln} shows that
the maximum voltage sensitivity occurs when the probe 
is close to the magnetic source. 
The conducting shell surrounding the magnet 
serves to shield the apparatus in the interior 
from external electric fields.  

In the presence of Lorentz violation,
rotating the entire apparatus 
produces a signal with a definite time variation,
which may increase sensitivity and reduce systematics.
The expected time variation of the signal
can be obtained by referring the laboratory coefficients
to the standard Sun-centered celestial-equatorial frame
\cite{kmphot},
which is an approximately inertial frame
appropriate for reporting results of 
arbitrary tests of Lorentz violation.
By virtue of the orbital speed 
$|\vec \be| \simeq 10^{-4}$ of the Earth in this frame,
a measurement of the three components of $\ko$ 
in the laboratory achieving a sensitivity $S$ 
then translates in the Sun-centered frame
into a sensitivity of order $S$ 
to the three components of $\ko$ 
and a sensitivity of order $10^4S$ to $\ktr$.

To illustrate these points,
consider the case of fixed probes 
recording a potential difference $\De\Ph$
between two points in the inner region.
We seek to characterize the expected time dependence 
of the signal due to the rotation of the Earth 
and its orbital motion about the Sun.
Other rotations,
such as those induced by turntables in the laboratory, 
can also be treated by these methods. 

In a frame fixed to the laboratory
and within the idealized approximations of the previous subsection,
$\De\Ph$ can be written as
\beq
\De\Ph =
({\cal M}_{DB})^{jk}_{\rm lab}(\ka_{DB})^{jk}_{\rm lab} 
= ({\cal M}_{DB})^{jk}_{\rm lab}(\ko)^{jk}_{\rm lab} ,
\label{dOlab}
\eeq
where $({\cal M}_{DB})_{\rm lab}$
is an experiment-specific constant matrix
that is determined for the chosen probe configuration
by applying Eq.\ \rf{finsoln}.
The time dependence of the signal $\De\Ph$
can be exhibited by transforming 
the laboratory-frame combinations $(\ko)^{jk}_{\rm lab}$ 
to the standard Sun-centered frame.
Following Ref.\ \cite{kmphot},
with upper-case letters denoting Sun-centered coordinates, 
we find
\bea 
(\ko)_{\rm lab}^{jk}
&=&
T_0^{jkJK}(\ka_{DB})^{JK}
\nonumber\\ &&
+(T_1^{kjJK} -T_1^{jkJK}) (\ka_{DE})^{JK} 
\nonumber\\ 
&=&T_0^{jkJK}(\ko)^{JK}
+2T_1^{kjJJ} \ktr 
\nonumber\\ &&
+(T_1^{kjJK} -T_1^{jkJK}) (\ke)^{JK} ,
\eea
where $T_0^{jkJK}=R^{jJ}R^{kK}$
and $T_1^{jkJK}= R^{jP}R^{kJ} \ep^{KPQ} \be^Q$
are tensors containing the time dependence
induced by the action of the rotations $R^{jJ}$
and boost $\be^J$.
Appendix C of Ref.\ \cite{kmphot}
provides explicit expressions for these quantities
and fixes the coordinate choices
for the laboratory and Sun-centered frames. 
 
In performing an experiment along the above lines,
some attention should be given to possible unconventional effects
induced by the apparatus,
in addition to the usual variety of systematic effects
such as surface patch charges. 
For example,
certain voltmeters are based on devices that measure currents.
The currents are determined 
by the dipole moment of a coiled wire,  
which is measured using an internal magnetic field 
to determine the torque.  
The presence of Lorentz violation implies 
this internal magnetic field could generate 
a corresponding electric field 
that could interfere with the signal 
from the magnetized sphere.  
In practice,
devices of this type could still be used under appropriate conditions,
such as an internal magnetic field significantly weaker 
than that of the magnetized sphere.   
As another caution,
inspection of the solution \rf{finsoln} shows that 
if the magnetic material chosen has a large dielectric constant
then the signal would be suppressed,
so a magnetic source of small dielectric constant is preferable.

Related experiments that could provide interesting sensitivity
to $\ko$ and $\ktr$ may also be possible.
For example,
a kind of converse of the above experiment could involve
attempting to measure a magnetic field created 
from a source of charge,
in analogy with Eq.\ \rf{d}.
Modern SQUID measurements of the magnetic field 
at the level of $10^{-14}$ T 
from a large vacuum electric-field source 
of $10^{12}$ V/m could in principle yield
comparable bounds to those above.

Another approach could be to take advantage of the
strong electric fields in the vicinity of an atomic nucleus.
Atomic spectroscopy might then reveal
the small accompanying Lorentz-violating magnetic field
through observable frequency shifts.
For this case,
we can adapt some existing theoretical and experimental studies 
of Lorentz-violating effects in atoms
\cite{ccexpt}.
Suppose as before the 10 birefringence-inducing coefficients 
in the photon sector are negligible.
A coordinate transformation of the type described in the appendix
can then be used to move the remaining 9 coefficients 
in the photon sector to the matter sector,
where they appear as symmetric components
of a $c$-type coefficient for Lorentz violation, 
$c_{\mu\nu} \supset \kft\al\mu\nu$.
For example,
the parity-odd coefficients $(\ko)^{jk}$ of interest above 
are contained in the three symmetric combinations $(c_{0j}+c_{j0})$.
To date no clock-comparison experiment 
has measured parity-odd $c$-type coefficients,
but future laboratory or space-based experiments 
could achieve Planck-scale sensitivities 
\cite{spaceexpt}
by incorporating into the analysis the boost effects 
arising from the orbital motion of the Earth or a satellite.

\acknowledgments
This work was supported in part
by the United States Department of Energy
under grant DE-FG02-91ER40661
and by the National Aeronautics and Space Administration
under grants NAG8-1770 and NAG3-2194.

\appendix
\label{appendix}

\section{Classical point charge}
\label{Classical point charge}

This appendix discusses some issues associated with
the choice of coordinate system in the presence of 
Lorentz violation
\cite{fn2}.
For definiteness,
we consider the theory of a single classical charged particle
in Lorentz-violating electrodynamics. 

The action is taken to be  
\bea 
S &=&  S_0 + S_{\rm int} + S_{\rm em} , 
\label{action1}
\eea
where $S_0$ is the action for the free classical particle,
$S_{\rm int}$ contains the interaction, 
and $S_{\rm em}$ is the action 
containing the pure-photon part of the lagrangian \rf{L_em1}.
The free action $S_0$ is 
\beq
S_0
= -m \int d\la 
\sqrt{\fr {dx^\mu}{d\la} \fr{dx^\nu} {d\la} \et_{\mu\nu}} ,
\label{S_matter} 
\eeq
and the interaction is assumed conventional,
\beq
S_{\rm int} = - \int d^4x ~ j^\mu A_\mu 
= -q \int d\la ~ A_{\mu}(x^\al) 
\fr{dx^\mu}{d\la}.
\label{S_int}
\eeq
As usual,
the equations of motion for the charged particle 
are obtained by varying with respect to $x^\al (\la)$ 
and reparametrizing with the proper time 
$d\ta^2 = \et_{\mu\nu}dx^\mu dx^\nu$.  
This shows that the conventional Lorentz-force law,
\beq 
m \fr{d^2 x^\mu}{d\ta^2} = 
q F^\mu_{\pt{\mu}\al} \fr{d x^\al}{d\ta},
\label{eqmotchg1}
\eeq  
holds despite the Lorentz violation
in the photon sector.

A suitable coordinate transformation can move
9 of the 19 coefficients for Lorentz violation 
from the photon sector to the matter sector,
while leaving unaffected the form $j^\mu A_\mu$ 
of the interaction.  
For simplicity in what follows,
we keep only the relevant 9 coefficients,
which corresponds to restricting the pure-photon lagrangian 
to Eq.\ \rf{L_em2},
and we work at leading order in the coefficients $\kfi$
for Lorentz violation.
The relevant coordinate transformation is
$x^\mu \to x^{\mu '} 
= x^\mu - \half (\kf)^{\al\mu}_{\pt{\al\mu}\al\nu} x^\nu$.
Under this transformation,
the total action becomes 
\bea 
S &=& \int d^4x ~ (-\frac 1 4 F_{\mu\nu}F^{\mu\nu} - j^\mu A_\mu )
\nonumber\\ & & 
- m \int d\la \sqrt{{dx^{\mu} \over d\la} {dx^{\nu} \over d\la} 
(\et_{\mu\nu} + \kft\al\mu\nu)}~.
\label{action2}
\eea

In the new coordinates,
the electromagnetic field has conventional 
kinetic and interaction terms and thus conventional dynamics.  
However, 
the charged particle now has Lorentz-violating dynamics 
controlled by the coefficients $\kft\al\mu\nu$.
These are the classical analogue of 
the symmetric traceless coefficients $c_{\mu\nu}$
in the matter sector of Lorentz-violating quantum electrodynamics.
The equations of motion for the classical charged particle 
are now expressed in terms of the proper time 
$d\ta^2 = (\et_{\mu\nu} + \kft\al\mu\nu) dx^\mu dx^\nu$,
and they represent a modified Lorentz force,
\bea
m \fr{d^2 x^\mu}{d\ta^2} 
+ m (\kf)^{\al\mu}_{\pt{\al\mu}\al\be} \fr{d^2 x^\be}{d\ta^2}
= q F^\mu_{\pt{\mu}\al} \fr{dx^\al}{d\ta}.
\label{eqmotchg2}
\eea  

Phenomenological analysis could proceed with either 
of the two actions \rf{action1} or \rf{action2},
or indeed with various other actions obtained by coordinate
transformations of Eq.\ \rf{action1}.
The physically observable effects are always equivalent,
but care is required in matching calculations to physical situations.
For example,
it might seem tempting to conclude that the action \rf{action1}
cannot describe physical Lorentz violation 
involving the 9 coefficients considered above
because there exists a coordinate system 
with conventional photon dynamics.
However, 
in practice charged particles are used
to measure properties of the electromagnetic field, 
and so in the new coordinate system the physical Lorentz violation
appears because observables are affected 
by the modified force law \rf{eqmotchg2}.
Determining which set of coordinates is most appropriate 
for a given experiment
involves establishing the underlying choice of  
standard rods and clocks to which the experimental observables
are ultimately being referenced. 

Comparisons between results obtained 
with the two different actions must include the appropriate 
coordinate transformation.
For example, observers 
using the two different actions \rf{action1} and \rf{action2} 
disagree on the force between two charged particles.
Consider an observer for whom two point charges $q$ and $q'$ 
are at rest and separated by a distance $\vec x$.
If the observer uses the action \rf{action1},
then the force between the two point charges is obtained
using the Lorentz-violating result \rf{c} 
together with the conventional Lorentz force \rf{eqmotchg1}.  
At leading order,
the force on charge $q'$ due to charge $q$ is then
\bea
m\fr {d u^0}{d\ta} &=& 0, 
\quad
m\fr {d u^j}{d\ta} 
=
q' E^j(\vec x),
\nonumber\\  
\label{fourforce1} 
\eea
where $E^j(\vec x)$ is the electric field \rf{d}
of the charge $q$. 
A second observer using the action \rf{action2}
finds instead a force given by applying the corresponding
coordinate transformation to Eq.\ \rf{fourforce1},
\bea
m \fr{d u^{0'}}{d\ta} &=& 
\fr{-q q'} {8\pi |\vec x'|^2}
(\kf)^{0j'l'j'} \hat x^{l'} ,   
\nonumber\\
m\fr {d u^{j'}}{d\ta}
&=& 
\fr {q q'}{4\pi |\vec x'|^2}
\big( \hat x^{j'} + 2(\kf)^{0j'0k'} \hat x^{k'} 
- \half (\kf)^{0'l'j'l'} \tilde x^{0'} 
\nonumber\\ & & 
\qquad\qquad
+ \frac 3 2 (\kf)^{0'l'k'l'} \hat x^{j'} \hat x^{k'} \tilde x^{0'} 
\big),
\label{fourforce2}
\eea
where $\tilde x^{0'}= x^{0'}/|\vec x'|$.
Note that the charges are moving in this second frame.
The result \rf{fourforce2}
can be derived directly 
from the modified Lorentz force \rf{eqmotchg2} 
with conventional photon dynamics,
provided care is taken to account for the
motion of the charges.

\end{document}